\documentclass[aps,prx,reprint,superscriptaddress,twocolumn,showkeys,amsmath,amssymb,longbibliography,floatfix,tikz]{revtex4-2}
\usepackage[english]{babel}
\usepackage{amsmath,amssymb,bbm,mathrsfs,bm,braket,color,graphicx,comment,amsfonts,dsfont}
\usepackage{hyperref}
\hypersetup{colorlinks=true,urlcolor=blue,linkcolor=blue,citecolor=blue}
\usepackage[mathscr]{euscript}
\usepackage{bm}
\usepackage{physics}

\newcommand{\ctwo}{\mathcal{C}_2 }
\newcommand{\trs}{\mathcal{T} }
\newcommand{\ct}{\mathcal{C}_2\mathcal{T} }

\begin{document}

\title{Robust Signatures of Fragile Topology in Nonmagnetic Materials}

\author{Viktor K{\"o}nye}
\thanks{These two authors contributed equally}
\affiliation{\mbox{Institute for Theoretical Solid State Physics, IFW Dresden, 01069 Dresden, Germany}}
\affiliation{\mbox{W\"urzburg-Dresden Cluster of Excellence ctd.qmat, Germany}}

\author{Ihor Nimyi}
\thanks{These two authors contributed equally}
\affiliation{\mbox{Institute for Theoretical Solid State Physics, IFW Dresden, 01069 Dresden, Germany}}
\affiliation{\mbox{W\"urzburg-Dresden Cluster of Excellence ctd.qmat, Germany}}
\affiliation{Kyiv Academic University, 03142 Kyiv, Ukraine}

\author{Oleg Janson}
\affiliation{\mbox{Institute for Theoretical Solid State Physics, IFW Dresden, 01069 Dresden, Germany}}

\author{Jeroen van den Brink}
\affiliation{\mbox{Institute for Theoretical Solid State Physics, IFW Dresden, 01069 Dresden, Germany}}
\affiliation{\mbox{W\"urzburg-Dresden Cluster of Excellence ctd.qmat, Germany}}
\affiliation{Institute  for  Theoretical  Physics,  TU  Dresden,  01069  Dresden,  Germany}

\author{Jasper van Wezel}
\email{vanwezel@uva.nl}
\affiliation{Institute for Theoretical Physics Amsterdam, University of Amsterdam, Science Park 904, 1098 XH Amsterdam, The Netherlands}

\author{Cosma Fulga}
\affiliation{\mbox{Institute for Theoretical Solid State Physics, IFW Dresden, 01069 Dresden, Germany}}
\affiliation{\mbox{W\"urzburg-Dresden Cluster of Excellence ctd.qmat, Germany}}

\date{\today}
\begin{abstract}
Some topological phases of matter are called fragile.
They can be trivialized by adding additional valence bands and do not generically host protected gapless boundary states.
Here we establish a fragile topological index with a bulk signature: it guarantees the presence of gap closing points in the bulk band structure of systems with both time-reversal and twofold rotation symmetry.
This result extends the established use of the Euler class in diagnosing the presence of Dirac cones for general two-dimensional systems with a combination of time-reversal and twofold rotation symmetry, to the case of nonmagnetic materials obeying both symmetries individually. The abundance of natural nonmagnetic materials with these symmetries renders the topological invariant practically useful, despite its fragility. We demonstrate this by applying first-principles calculations to five materials and finding that all of them host Dirac points identified by the invariant, suggesting its widespread applicability.
Our results thus provide a robust spectroscopic signature of fragile topology in nonmagnetic systems which can be directly accessed in experiments on practically relevant materials.
Furthermore, they enable an alternate method for finding Dirac cones in ab-initio simulations, and may be used as a route towards identifying materials with nonlinear transport properties.
\end{abstract}

\maketitle

\section{Introduction}
Topological phases of matter are commonly associated with robust gapless boundary states \cite{Hasan_2010, Qi_2011}.
These cannot be removed unless either the bulk gap closes at the Fermi level, or the protecting crystal symmetries are broken.
Moreover, their presence is independent both of the total number of bands in the system, and of any band crossings that may occur deep in the valence or conduction sectors.
Topological phases which obey these properties are said to be \emph{stable} \cite{Kitaev_2009, Brouwer_2023}.

Not all topological phases are stable.
A prominent exception is provided by so-called \emph{fragile} phases \cite{Po_2018, Song_2020, Bouhon_2020, Lee_2025}, which are not generically associated with gapless boundary states, and whose topology depends on the total number of valence bands, even far away from the Fermi level.
For a given number of bands, such phases are topological in the sense that it is not possible to construct exponentially-localized and symmetric Wannier functions characterizing the valence band.
However, if another valence band is added to the system, a Wannier representation may become possible, in which case one says that the phase has been trivialized.

Experimentally measuring the signatures of fragile topology poses significant challenges.
One of the hallmark features of a fragile phase is the presence of boundary charges localized either on corners or on hinges \cite{Benalcazar_2019, Wieder_2020, Kooi_2021}, which could be obscured in real materials by local impurities close to the surface, or could be screened in the case of a metallic system.
An alternate signature corresponds to a gap closing that occurs in the presence of twisted-periodic boundary conditions \cite{Song_2020a}.
In the case of twisted bilayer graphene \cite{Bistritzer_2011, Cao_2018, Cao_2018a}, fragile topology has been theoretically predicted to lead to a crossing between the Landau levels originating from the first and second valence bands \cite{Lian_2020, Guan_2022}.
Finally, in three-dimensional systems with only two occupied bands, fragile topology may lead to the presence of gapless surface or hinge states, but the latter can be gapped out by adding additional occupied states \cite{Wieder_2020, Kobayashi_2021, Sato_2025, Tanaka_2026}.
In light of the difficulties associated with its experimental demonstration, direct observations of fragile topology have so far only been reported in highly-controllable metamaterial platforms \cite{Peri_2020, Zhao_2022, Jiang_2024, Wu_2024}.

A bulk signature of fragile topology has been proposed for two-dimensional (2D) systems with combined time-reversal ($\trs$) and twofold rotation ($\ctwo$) symmetries. In these, the Euler class characterizing the topology also provides a lower bound for the number of Dirac points in the energy bands \cite{Ahn_2019, Bouhon_2020}. In this way, fragile topology obtains a robust spectroscopic signature.

Here, we extend this result to the class of nonmagnetic systems obeying $\trs$ and $\ctwo$ individually. Although this gives a more stringent condition than the original case, it is of direct practical relevance, as there are significantly more nonmagnetic (i.e. $\trs$-preserving) than magnetic crystals in existing databases \cite{Itani_2025}. Moreover, an extension of the original Euler class argument is needed for this case, since in systems with $\trs$, Kramers' theorem enforces the presence of Dirac points at time-reversal-invariant momenta (TRIM). In order to give a lower bound for the number of Dirac points present at generic points of the Brillouin zone (BZ), the contribution from TRIM needs to be separated.
To this end, we establish a relation between two existing fragile topological indexes, one obtained from the Wilson loop spectrum \cite{Bradlyn_2019, Bouhon_2019, Henke_2021, Li_2026}, and the other from the Euler loop spectrum \cite{Unal_2020, Ahn_2019}.
If these indexes do not match (i.e.~their difference is nonzero), then there must exist generic band crossing points in the the BZ next to the Kramers' degeneracies.
Their position in momentum space is not constrained by symmetry, so they can be located at arbitrary points in the BZ, but fragile topology guarantees their existence.

We demonstrate the practical applicability of our result by performing first-principles calculations on five materials from the Computational 2D Materials Database \cite{Haastrup_2018, Gjerding_2021}. We find that all of them host Dirac cones at generic momenta, suggesting that fragile topology may have widespread consequences in real materials.

Beyond identifying a direct signature of fragile topology in nonmagnetic materials that can be accessed using angle-resolved photoemission spectroscopy (ARPES) \cite{Sobota_2021}, our results also provide a method for finding Dirac points in first-principles calculations, similar to how Weyl points are found in 3D systems by computing Chern numbers \cite{Gresch_2017}.
Further, the fragile-topology-enforced Dirac cones will generically be gapped under perturbations breaking rotation symmetry (e.g.~a crystal field), leading to a nonzero Berry dipole \cite{Sodemann_2015}.
This suggests that fragile topology may be used as a route towards identifying materials with robust nonlinear transport properties.

\section{Results}
\label{sec:results}

We focus on 2D, spinful, many-band systems obeying $\ctwo$ and $\trs$ symmetries, where the twofold rotation is along an axis perpendicular to the plane of the material.
In the basis where the two operators commute, we have $\trs^2=\ctwo^2=-1$, meaning that their product, $(\ct)^2=+1$.
The latter condition ensures that there always exists a basis in which the Hamiltonian is purely real at every momentum \cite{Ahn_2019}, $H({\bf k})=H^*({\bf k})$ with ${\bf k}=(k_x,k_y)$ the dimensionless momentum vector.
Further, time-reversal symmetry forces all states to be doubly-degenerate at all TRIM due to Kramers' theorem.
As such, the systems we consider will generically be described by pairs of bands, where the two bands in a pair touch at all four TRIM points, leading to the formation of four Dirac cones.
Crucially, however, $\ct$ symmetry also allows for pairs of protected Dirac cones to exist at generic momenta in the BZ \cite{Zhao_2017, Ahn_2019, Bouhon_2020}, which we label as \emph{generic Dirac cones} so as to distinguish them from TRIM cones.
It is these generic cones whose presence can be enforced by fragile topology, as we will discuss below.

In the following, we will focus on one pair of bands in a multi-band system. We assume that for certain parameter values, this pair may be decoupled from all others, meaning that there exists an energy gap at all momenta between this band pair and the ones at energies above and below it.
For purposes of illustration, we will use the four-band toy model introduced in Ref.~\cite{Henke_2021}. 
Its Bloch Hamiltonian is given by:
\begin{align}
\label{eq:Ham}
H(\vb{k}) &= \cos{k_x}\sigma_0\tau_x + \cos{k_y}\sigma_y\tau_y \notag \\ 
&\phantom{=} + \left[m^2 - (\sin{k_x}+\sin{k_y})^2\right]\sigma_0\tau_z \notag \\
&\phantom{=} - \frac{\sin{k_x}}{2}\sigma_x\tau_z +
\frac{\sin{(k_x+k_y)}}{10}\sigma_z\tau_z.
\end{align}
Here $\sigma_i$ and $\tau_i$ are Pauli matrices corresponding to the spin and orbital degrees of freedom, respectively, and $m>0$ is a real parameter. This Hamiltonian respects ${\cal C}_{2}H(\vb{k}){\cal C}_{2}^\dag = H(-\vb{k})$, and ${\cal T}H(\vb{k}){\cal T}^\dag = H(-\vb{k})$, where ${\cal C}_{2}=i\sigma_y\tau_0$ and ${\cal T}=-i\sigma_y\tau_0K$ (with $K$ denoting complex conjugation).
Note that this Hamiltonian has been rotated with respect to its original basis of Ref.~\cite{Henke_2021}, in which $\ctwo$ was proportional to $\sigma_z\tau_0$.
This has been done so that the combined symmetry takes the form $\mathcal{C}_{2}\mathcal{T}=K$, and $H$ is purely real.

\begin{figure*}[tb]
    \centering
    \includegraphics[width=1\textwidth]{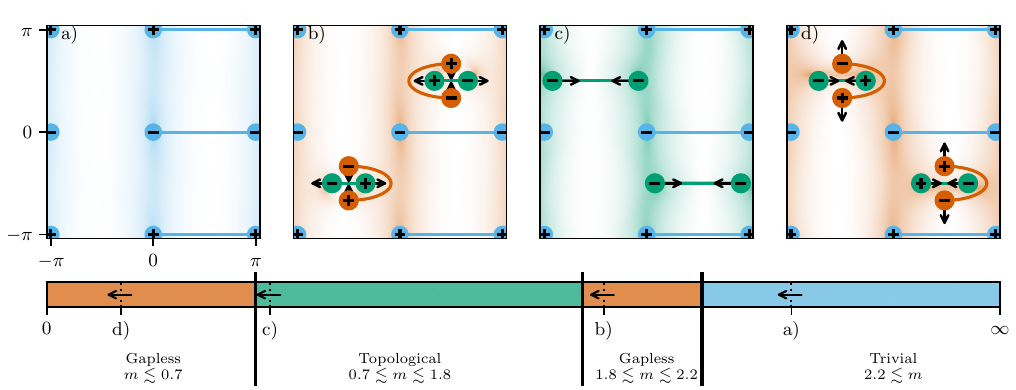}
    \caption{Topologically distinct phases of the Hamiltonian in Eq.~\eqref{eq:Ham} as a function of $m$, with transition points indicated along the bottom bar. 
    In the orange regions the gap at the Fermi energy is closed. 
    As the gap opens in the green (blue) region we encounter a topological (trivial) insulator with $\chi=-2$ ($\chi=0$). 
    The panels (a-d) show the degeneracies between various bands at specific values of the parameter $m$. The intensity of shading in each panel indicates the difference in energy between the two occupied bands on a logarithmic scale. 
    The signs $+$ and $-$ inside colored circles denote winding numbers $+1$ and $-1$ respectively. 
    The Dirac points at the TRIM and Dirac strings between them are marked in blue, while the Dirac points formed between the occupied bands but away from the TRIM are depicted in green, as are the Dirac strings connecting them. 
    The orange circles indicate Dirac points connecting valence and conduction (the second and third) bands, with orange Dirac strings linking them. 
    The arrows denote the direction of movement for all Dirac points as $m$ decreases. 
    }
    \label{fig:model_phases}
\end{figure*}

Using this toy model as a concrete example, we will describe the generic Dirac cones of a band pair by introducing two main quantities: Euler loops and Wilson loops.

\subsection{Euler loops}

Due to $\ct$, each Dirac point can be characterized by an integer winding number $w_i$ that can be calculated using the Euler connection.
The latter is defined as $\vb{a}(\vb{k})=\bra{u_{1}(\vb{k})}\nabla_k\ket{u_2(\vb{k})}$ \cite{Zhao_2017, Ahn_2019}, where $\ket{u_{1,2}(\vb{k})}$ are the Bloch states of a band pair, in the basis in which the Hamiltonian is real.
We will refer to the integral of the Euler connection over a closed loop $\ell$ as the Euler loop $\mathcal{E}_\ell$, defined by:
\begin{equation}
\label{eq:Eulerloop}
    \mathcal{E}_\ell = \frac{1}{\pi}\oint\limits_\ell \dd{\vb{k}}\cdot \vb{a}(\vb{k}).
\end{equation}
The Euler loop is not quantized in general, except in two-band systems. 
However, even in many-band systems, $\mathcal{E}_\ell$ will still converge to an integer value provided that the loop is infinitesimally small. 
This gives the winding number of the Dirac points: 
$w_i=\mathcal{E}_{\ell_i}=\pm 1$ , where $\ell_i$ is an infinitesimally-small loop surrounding a Dirac cone (for details see Ref.~\cite{Ahn_2019} and Appendix \ref{methods:Euler}). 

The sum of all Dirac cone winding numbers gives a fragile topological index characterizing an isolated band pair, which is called the Euler class, $\mathop{\chi}=\frac{1}{2}\sum_iw_i$~\cite{Ahn_2019}.
This index is conserved provided the band pair remains isolated, meaning that generic (i.e.~non-TRIM) Dirac cones can be created or annihilated pairwise only if they have opposite winding numbers.
However, if there is a gap closing and reopening between two adjacent band pairs, the sum of their Euler classes is no longer conserved and can be trivialized, as expected for a fragile index (see Appendix \ref{methods:fragileEuler} for an example).
In terms of winding numbers, this means that when two band pairs touch, the winding numbers of Dirac cones can change sign, via a process which is known as Dirac cone braiding \cite{Ahn_2019,Bouhon_2020a, Jiang_2021, Slager_2024, Breach_2024, Lee_2024}.

In Fig.~\ref{fig:model_phases}, we show examples of Dirac cone configurations for different values of the parameter $m$ in the toy model of Eq.~\eqref{eq:Ham}. 
The blue circles indicate the TRIM cones of the pair of valence bands (which we refer to as the first and second band), and each cone is labeled with its winding number.
The green circles show generic cones in the same valence band pair, while generic cones connecting the highest valence and the lowest conductance band (the second and third bands) are marked by orange circles. 
Because of the symmetries of the system, the degeneracies of the pair of conduction bands (not shown) will be at the same momenta as those of the valence bands. 
The figure also shows Dirac strings (discontinuities in the gauge, across which the eigenvectors change sign \cite{Ahn_2019}).
The latter connect pairs of degeneracies, and are represented in a particular gauge choice.

For $m\gtrsim 2.2$, shown in Fig~\ref{fig:model_phases}(a), there are no degeneracies except at the TRIM.
Lowering the value of $m$, the gap between the conduction and valence bands closes and the two pairs of orange Dirac points appear with opposite winding numbers.
Additionally, two pairs of green Dirac points  with opposite winding number appear in the occupied bands, as shown in Fig~\ref{fig:model_phases}(b).
Further decreasing the value of $m$, the orange Dirac points annihilate in pairs. 
In doing so, however, their Dirac strings each cross one of the green points, whose winding number then changes. 
This is what is referred to as Dirac cone braiding.
The orange point also needs to cross a green Dirac string, but this does not change its winding number because there is a second green string in the conduction bands (not shown) that is crossed at the same time. 

After the orange Dirac points are annihilated, as shown in Fig~\ref{fig:model_phases}(c), the gap between valence and conduction bands is open again at all momenta, and we are left with two pairs of green Dirac points within the valence bands that all have the same winding numbers. 
At even lower values of $m$, the gap between valence and conduction bands closes once again, and the same braiding process repeats [see Fig~\ref{fig:model_phases}(d)], allowing the cones to be annihilated .

Thus, as a function of $m$, the system exhibits two insulating phases separated by a gapless regime.
For $m\gtrsim 2.2$ (blue region), there are no generic Dirac cones and the sum of all winding numbers in the valence band pair is $\chi=0$.
For $0.7 \lesssim m \lesssim 1.8$ (green region), $\chi=-2$ and there are generic Dirac cones which all have the same winding number, meaning that they cannot be pairwise annihilated.
Our aim is to show that fragile topology \emph{enforces} the presence of these Dirac cones, which requires us to introduce a second fragile index.

\subsection{Concentric Wilson loops}

Next to the Euler loop we also use the Wilson loop (WL) to characterize an isolated band pair.
Using the non-Abelian Berry connection matrix $\mathbf{A}_{mn}(\mathbf{k}) = i\mel{u_m(\mathbf{k})}{\nabla_k}{u_n(\mathbf{k})}$ \cite{Yu_2011} 
the $2\times 2$ WL matrix is defined on a closed loop ${\ell}$ as:
\begin{equation}
\label{eq:wilson}
{\cal W}_\ell = {\cal P}\text{ exp}\left(i\oint_{\ell}d\mathbf{k}\cdot\mathbf{A}\right).
\end{equation}
Here ${\cal P}$ is the path ordering operator (for details see Appendix \ref{methods:CWLS}).
Note that as long as the loop is closed, ${\cal W}_\ell$ is unitary, and its two eigenvalues are complex phases.

\begin{figure*}[tb]
    \centering
    \includegraphics[width=0.9\textwidth]{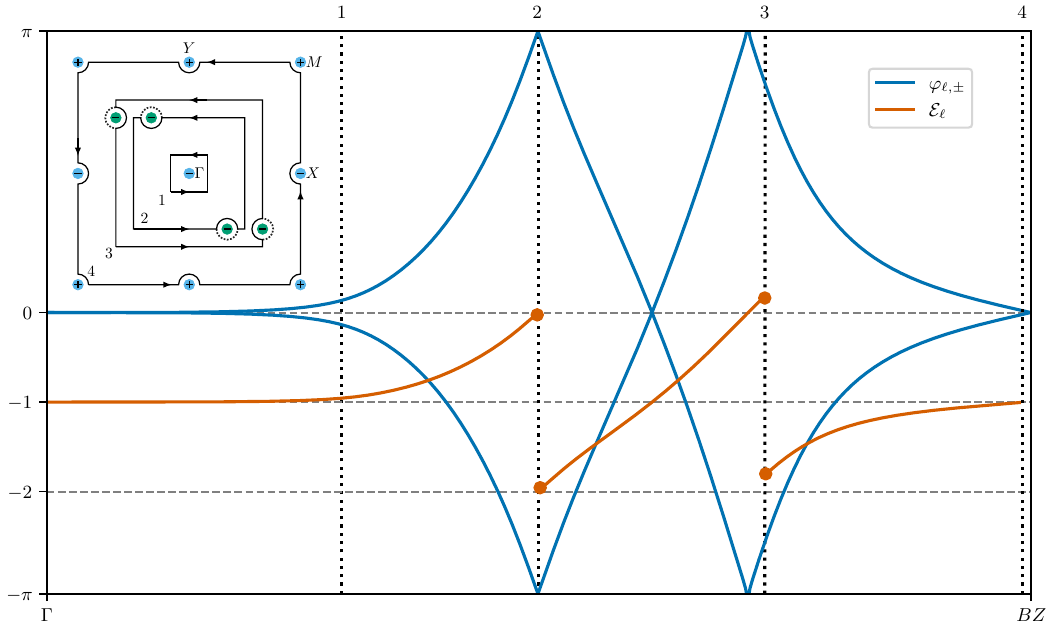}
    \caption{The spectrum of concentric square Wilson loops (blue) and the corresponding Euler loop (orange) for the Hamiltonian in Eq.~\eqref{eq:Ham} with $m=1.75$. The horizontal axis represents the radius of the WL, which starts out tightly enclosing the $\Gamma$ point, and eventually covers the entire BZ. The vertical black dotted lines labeled 1-4 indicate the radii of the corresponding numbered Wilson loops shown in the inset.
    }
    \label{fig:eulersjumps_and_WLS}
\end{figure*}

We consider a set of loops that are concentric squares around the $\Gamma$ point of the BZ with increasing radii, starting from one which is infinitesimally close to $\Gamma$ and ending with one that encircles the entire BZ.
Due to $\ct$, the two eigenphases of ${\cal W}_\ell$ are always opposite, $\varphi_{\ell, +} = -\varphi_{\ell, -}$ for each such loop $\ell$.
Taken together, the evolution of $\varphi_{\ell, \pm}$ as a function of loop radius gives the concentric Wilson loop spectrum (CWLS), which can wind as a function of loop radius.
This winding number, $w_{\text{CWLS}}$, has been considered before as an indicator of fragile topology \cite{Bradlyn_2019, Bouhon_2019, Li_2026}, and also discussed in Ref.~\cite{Henke_2021}.
Using the toy model of Eq.~\eqref{eq:Ham}, we find that for $m=1.75$ each of the WL eigenphases of the valence band pair winds twice as a function of loop radius (blue line in Fig.~\ref{fig:eulersjumps_and_WLS}), yielding $w_{\text{CWLS}}=2$.
Note that the two eigenphases always wind in opposite directions, so we use a convention in which $w_{\text{CWLS}}$ is positive.
This winding number cannot change provided that the band pair remains isolated, since the $\pi$ (and the $0$) crossings of the CWLS are protected by $\ct$ \cite{Bouhon_2019}.

\subsection{Wilson-Euler equivalence}

In Appendix \ref{methods:WilsonEuler} we show that within a set of concentric loops $\ell$, the Euler loop and the eigenvalues of the Wilson loop for an isolated band pair are related to each other by:
\begin{equation}
\label{eq:wilsoneig}
    \varphi_{\ell,\pm} = \pi(1 \pm \mathcal{E}_\ell) \mod 2\pi.
\end{equation}

This is our main technical result.
We illustrate this relation in Fig.~\ref{fig:eulersjumps_and_WLS}, which shows the evolution of the CWLS and of the Euler loops (blue and orange lines respectively) as a function of radius.
The smallest numbered loop shown in the inset of Fig.~\ref{fig:eulersjumps_and_WLS} closely encircles $\Gamma$, so the value of its Euler loop (orange) is close to the winding number associated with the $\Gamma$ point, which in this example is $\mathcal{E}_{\ell_\Gamma} \simeq w_\Gamma = -1$.
As the concentric loops in the BZ grow, the values of their Euler loops initially change continuously. Since the concentric loops are taken such that they are $C_2$ symmetric, however, they will encounter pairs of symmetry-related Dirac points as they expand.
The loops just before and just after crossing a pair of Dirac points differ by only the contributions coming from the Dirac points themselves (as indicated schematically by semicircles around the Dirac points in the inset of Fig.~\ref{fig:eulersjumps_and_WLS}).
The difference between the corresponding Euler loop values will thus be the winding number of the two Dirac points, yielding a jump of $-2$ in the Euler loop spectrum whenever a pair of bulk Dirac points is crossed.
The final loop reaches the edge of the BZ, at which point the straight sections of the loop cancel one another via the periodicity of the BZ, and all that remains are small loops closely encircling the TRIM in a clockwise direction. For the current example, the Euler loop value at the maximal radius approaches $\mathcal{E}_{\ell_{BZ}}=-\left(w_X+w_Y+w_M\right)=-1$, with the minus sign coming from the clockwise orientation of the small loops.

In this particular example, the concentric Euler loops at $\Gamma$ and at the edge of the BZ have equal values, but that is not always the case.
From the previous formulas, however, we can see that the difference between these two loops is always equal to $\mathcal{E}_{\ell_\Gamma}-\mathcal{E}_{\ell_{BZ}}=w_\Gamma+w_X+w_Y+w_M$ (neglecting the possibility of additional Dirac points being present on the BZ boundary).

In general, for a pair of isolated bands, the concentric Euler loop as a function of loop radius is a curve that (i) starts and ends at an odd integer, (ii) undergoes jumps of $\pm 2$ each time the loop crosses a pair of generic Dirac points with winding number $\pm 1$, and (iii) is continuous otherwise.
We can also see both in Fig.~\ref{fig:eulersjumps_and_WLS} and directly from Eq.~\eqref{eq:wilsoneig} that every time the value of an Euler loop is an even integer, the eigenphases of the associated WL equal $\pi$. 
In the case shown in Fig.~\ref{fig:eulersjumps_and_WLS} there are two such $\pi$ crossings of the CWLS, and both are protected since $w_\text{CWLS}=2$.
Finally, note that Eq.~\eqref{eq:wilsoneig} is fully consistent with the CWLS being continuous in spite of the Euler loop jumping by an even integer.

\begin{figure*}[tb]
    \centering
    \includegraphics[width=0.85\linewidth]{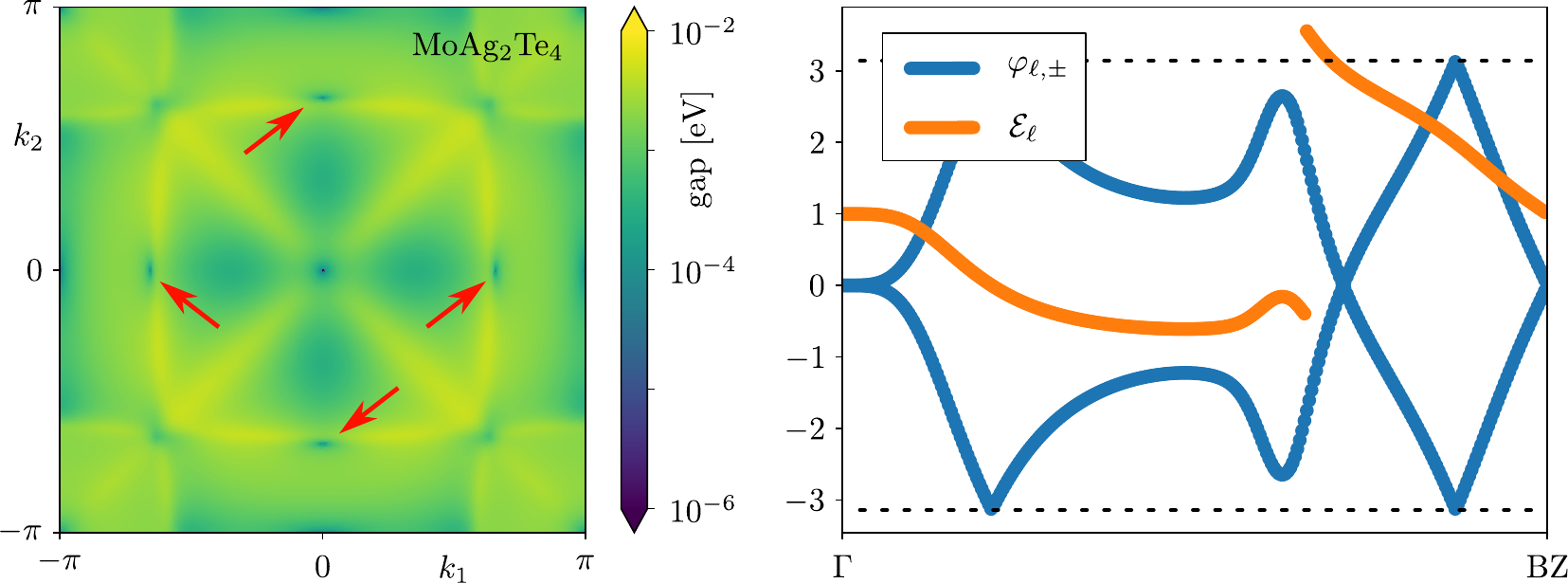}
    \caption{Dirac cones enforced by fragile topology in MoAg$_2$Te$_4$.
    The left panel shows the energy gap between the two bands in a pair, with generic Dirac cones (all winding numbers equal to $+1$) indicated by red arrows.
    $k_{1,2}$ are dimensionless momenta in the directions of the two Bravais vectors of the lattice.
    The right panel is the same as as Fig.~\ref{fig:eulersjumps_and_WLS}, showing the Euler loop and the CWLS.
    This is the third band pair in order of increasing energies, formed by bands numbers five and six. As such, it is located deep in the valence sector around $-10$ eV to $-11$ eV.}
    \label{fig:MoAg2Te4}
\end{figure*}

\subsection{Fragile topological index enforcing Dirac cones}

Taken together, the winding number of the CWLS and the winding numbers of the TRIM cones can be used to predict that the band pair must contain generic Dirac cones.
This can be shown by contradiction.

Assume that there do not exist any generic Dirac cones in a band pair. 
Then $\mathcal{E}_{\ell}$ is a continuous function of loop radius.
It starts at an odd value $\mathcal{E}_{\ell_\Gamma} = w_\Gamma$ and ends at an odd value $\mathcal{E}_{\ell_{BZ}}=-\left(w_X+w_Y+w_M\right)$.
Thus, the minimum number of times that $\mathcal{E}_{\ell}$ becomes even is half of the difference between the end points:
$|\mathcal{E}_{\ell_\Gamma}-\mathcal{E}_{\ell_{BZ}}|/2=|w_\Gamma+w_X+w_Y+w_M|/2 = |\chi|$, where $|\cdot |$ denotes the absolute value.
Due to Eq.~\eqref{eq:wilsoneig}, this means that the CWLS must contain the same minimum number of $\pi$ crossings, i.e.~the same $w_\text{CWLS}=|\chi|$.
We express this requirement by defining the difference between these indexes,
\begin{equation}\label{eq:Q}
    Q = w_\text{CWLS}-\frac{1}{2} \left| \sum_{\text{TRIM}} w_i \right|,
\end{equation}
which should be equal to zero.

However, if the winding number of the CWLS does not match the sum of the winding numbers of the TRIM cones, meaning that $Q\neq 0$, then we reach a contradiction.
There are only two ways of resolving it.
First, it could be that there exist generic Dirac cones on the BZ boundary.
In this case, the Euler loop may be continuous, but the final value $\mathcal{E}_{\ell_{BZ}}$ is not given just by the sum of the winding numbers of the TRIM cones. The second option is that the Euler loop is not continuous, and thus contains jumps.
Each such jump is in general by $\pm 2$, corresponding to a pair of generic Dirac cones with winding number $\pm 1$.
With generic Dirac cones either at the BZ edge or in the BZ bulk, each pair of generic Dirac cones increases or decreases the minimum number of even values of $\mathcal{E}_{\ell}$ as a function of radius by one (either by means of a jump of $\pm 2$ or by changing the value of $\mathcal{E}_{\ell_{BZ}}$ by $\pm 2$).
In order to satisfy Eq.~\eqref{eq:wilsoneig}, there must then be \emph{at least} $2|Q|$ generic Dirac cones in the band pair.
This is only the minimum number, since it is always possible to pairwise create any additional number of generic Dirac cones with opposite winding numbers, without changing any of the terms of Eq.~\eqref{eq:Q}.

The relation Eq.~\eqref{eq:Q} can also be understood in a graphical way, using the toy-model example.
The Euler loop of Fig.~\ref{fig:eulersjumps_and_WLS} starts and ends at $-1$, but it must be even at least twice, because there are two protected $\pi$ crossings in the CWLS.
Therefore, the Euler loop cannot have a shape allowing it to be smoothly deformed to, say, being constant at $-1$ for all radii.
The solution is that the Euler loop must contain jumps, signaling the presence of generic Dirac cones.
In this example, the winding numbers of the TRIM cones sum to zero, and thus $Q=w_\text{CWLS}=2$ and there are four generic Dirac cones, as shown in Fig.~\ref{fig:model_phases}(b).

\subsection{Relation to the Euler class}

Before proceeding, we draw a distinction between our fragile index $Q$ and previous work examining the effect of fragile topology on Dirac cones.
As stated before, it is already known that the Euler class $\chi$ of a band pair equals the sum of Dirac cone winding numbers whenever the combined symmetry $\ct$ is preserved.
As such, a band pair with a nonzero Euler class necessarily must have some Dirac cones, either at TRIM or at generic BZ points. This statement was also explicitly made in Refs.~\cite{Ahn_2019, Bouhon_2020}.

However, when both $\trs$ and $\ctwo$ are present, it is the index $Q$ (and not the Euler class) that enables us to formulate a distinct spectroscopic signature of fragile topology, by distinguishing between TRIM cones and generic ones.
To see this, note that the Euler loop and the Wilson loop contribute to separate terms in the definition of $Q$ in Eq.~\eqref{eq:Q}, and both are needed.
For example, if there are no generic cones and only TRIM cones are present, then the band pair could have any value of $\chi \in \{ 0, \pm 1, \pm 2 \}$, but this difference in Euler class would not lead to any distinct features in the band structure.
Similarly, if there exist, say, four generic Dirac cones with winding number $+1$, while at the same time all four TRIM cones have winding numbers $-1$, then the Euler class vanishes.
However, such a band pair is not trivial:
It hosts generic Dirac cones enforced by fragile topology, which are predicted by a value of $Q\neq 0$.
We show an explicit example of this in a band pair of the material MoOBrCl in Appendix \ref{methods:materials}.

\subsection{First-principles calculations}

None of the above discussion relied on fixing a particular number of bands in the system.
As such, we expect that Eq.~\eqref{eq:Q} can be used to characterize the isolated band pairs of real materials. We have tested this prediction by considering 2D materials from the Computational 2D Materials Database (C2DB)~\cite{Haastrup_2018, Gjerding_2021}. For our proof-of-principle calculations, we selected five materials that include dynamically and energetically stable as well as unstable compounds: Au$_2$SO$_4$~\cite{Au2SO4}, Au$_3$InO$_4$Br$_4$~\cite{Au3InO4Br4}, MoOBrCl~\cite{MoOBrCl}, MoAg$_2$Te$_4$~\cite{MoAg2Te4}, and PdIr$_3$S$_4$Br$_4$~\cite{PdIr3S4Br4}.
See Appendix \ref{methods:DFT} for details of
density-functional-theory calculations and recasting the band structures into a
Wannier basis.

All materials were chosen such that they obey $\ctwo$ and $\trs$, and such that they break sufficiently many lattice symmetries in order to avoid band degeneracies along entire lines in the BZ, or across the entire BZ itself (i.e. no inversion).
We also ensured that there already existed band structure calculations included in the database, and we visually checked that they appear to have some isolated band pairs.
We remark that these are the only five materials that we have tested.

In each case, we rotated the Wannier Hamiltonian into its real basis and computed $Q$ for all isolated band pairs.
To confirm the presence of generic Dirac cones, we additionally computed the concentric Euler loops for all $Q\neq 0$ pairs.
We found that each of these materials hosts between three and six pairs of bands for which $Q\neq 0$, hinting at the possibility that Dirac cones enforced by fragile topology are ubiquitous in real crystals.
We show in Fig.~\ref{fig:MoAg2Te4} an example in the case of MoAg$_2$Te$_4$, which looks similar to Fig.~\ref{fig:eulersjumps_and_WLS} in the sense that the winding numbers of the TRIM cones sum to zero and $w_\text{CWLS}=Q=2$.
In this case, the four generic Dirac cones are related to each other by a fourfold roto-inversion symmetry, such that they are all positioned at the same loop radius and the Euler loop jumps by $+4$.
In Appendix \ref{methods:materials}, we show results also for the other materials, some of which have $Q\neq 0$ band pairs closer than 2 eV from the Fermi level, making them amenable to ARPES detection.

\section{Discussion}

In two dimensions with both ${\cal C}_{2}$ and ${\cal T}$ symmetry, the Dirac points of an isolated band pair are stable in the sense that they can be removed only by pairwise annihilation. 
This includes the possibility of two Dirac points of equal winding number coalescing with a TRIM cone of opposite winding number, which results in a TRIM cone of flipped sign and conserved total winding number. 
By relating two quantities, the Wilson loop and the Euler loop, we identify a fragile topological index $Q$ which, for an isolated pair of bands, enforces the presence of Dirac cones at arbitrary momenta in the BZ.
This index is independent of the total number of bands in the system, and appears to take nonzero values in a wide range of realistic materials ---in our first-principles calculations, five out of five crystals.
On the one hand, this index provides a direct signature of fragile topology that can be accessed in ARPES measurements.
Beyond this, it may allow an automated search for bulk Dirac cones in ab-initio codes, without requiring energy minimization, and it is also applicable to the 2D $\trs$- and $\ctwo$-invariant planes of a 3D BZ.
In the latter case, the generic ``Dirac cones'' on that 2D plane would correspond either to Weyl cones or to nodal lines in the full 3D BZ.

Finally, the same symmetry constraints $\ctwo$ and $\trs$ also ensure that the Berry curvature is zero throughout the BZ. 
Introducing a perturbation that breaks $C_{2}$, leaving only $\cal{T}$, will generically lift any degeneracies at bulk Dirac points, leading to non-zero Berry curvature $\Omega(\vb{k})$ concentrated close to the original location of the Dirac points. Since time-reversal symmetry enforces $\Omega(\vb{k})=-\Omega(-\vb{k})$, the anomalous Hall conductivity related to the Berry curvature created this way will vanish. However, a non-zero time-reversal symmetric Berry curvature generically yields a non-linear response related to its the Berry curvature dipole $\vb{D}$, defined as $D_i = \int_{\text{BZ}} \dd[2]{k} f(\vb{k})\partial_{k_i}\Omega(\vb{k})$
where $f(\vb{k})=1/(\mathrm{e}^{(E(\vb{k})-\mu n)/k_BT}+1)$ is the Fermi-Dirac distribution at temperature $T$, chemical potential $\mu$, and filling fraction $n$~\cite{Sodemann_2015}.

\begin{acknowledgments}
We thank Klaus K\"{o}pernik for fruitful discussions and Ulrike Nitzsche for technical assistance.
This work was supported by the Deutsche Forschungsgemeinschaft (DFG, German Research Foundation) under Germany’s Excellence Strategy through the W\"{u}rzburg-Dresden Cluster of Excellence on Complexity and Topology in Quantum Matter – ctd.qmat (EXC 2147, 390858490 and 392019).
\end{acknowledgments}

\section*{Data availability}

The data generated as part of this work is available upon request, or can be re-generated using the code available at \cite{Zenodo_code}.

\section*{Code availability}

The code used to generate our results uses the kwant package \cite{Groth_2014} and is available at \cite{Zenodo_code}.

\appendix

\section{Characterization of Dirac points via the Euler loop}
\label{methods:Euler}

The fact that the Hamiltonian is real implies that Dirac points (degeneracy points with linear dispersion) cannot be gapped out by symmetry-preserving perturbations. 
One way to see this for an isolated pair of bands is to notice that the $i$-th Dirac point is characterized by an integer winding number $w_i=\pm 1$, while any point in a single gapped band has vanishing winding number~\cite{Ahn_2019}. 
For a two-band system, the sum of these winding numbers give the Euler class $\mathop{\chi}=\frac{1}{2}\sum_iw_i$~\cite{Ahn_2019}.

More generally, for subsystems with a pair of bands that are well separated from other bands, the eigenvectors form a rank-2 real vector bundle on the 2D Brillouin Zone (BZ). The Euler class characterizing this vector bundle can then also be constructed from the Euler connection, which is defined as $\vb{a}(\vb{k})=\bra{u_{1}(\vb{k})}\nabla_k\ket{u_2(\vb{k})}$ \cite{Zhao_2017, Ahn_2019, Bouhon_2020}.

Since $\braket{u_n(\vb{k})}{u_m(\vb{k})}=\delta_{nm}$, the $2\times 2$ Berry connection matrix characterizing a band pair respects $\langle u_m(\vb{k})|\nabla_k|u_n(\vb{k}) \rangle = -\langle u_n(\vb{k})|\nabla_k|u_m(\vb{k}) \rangle^*$. 
As a consequence, and using real eigenvectors, the non-Abelian Berry connection for two sub-bands of a larger system therefore has to take the form:
\begin{equation}
\label{eq:connection}
    \vb{A} = \begin{pmatrix}
      0 & i\vb{a}(\vb{k}) \\
-i\vb{a}(\vb{k}) & 0  
    \end{pmatrix} = -\vb{a}(\vb{k})\sigma_y.
\end{equation}
Here $a(\vb{k})\in\mathbb{R}$ is the Euler connection \cite{Bouhon_2020}.
The reality of the eigenvectors is maintained by real diagonal gauge transformations of the form
\begin{equation}
\label{eq:realgaugetraf}
\ket{u_m(\vb{k})}'=(-1)^{\xi_m(\vb{k})}\ket{u_m(\vb{k})},
\end{equation}
where $\xi_m(\vb{k})\in\mathbb{Z}$ is an integer function. 
It can be shown that the Euler connection is invariant under gauge transformations of this type up to a global sign [i.e.~$\vb{a}'(\vb{k})=\pm\vb{a}(\vb{k})$ for transformations given by Eq.~\eqref{eq:realgaugetraf}]. This sign ambiguity can be removed by fixing the eigenstates for a given momentum and fixing the gauge continuously starting from that point. 

Using the fact that at non-degenerate momenta, it is always possible to locally choose a gauge with well-defined $\vb{a}(\vb{k})$ and continuous eigenstates, we can cover the whole BZ by patches with different local gauge choices, each of which contains a locally smooth Euler connection that is invariant under transformations of the type of Eq.~\eqref{eq:realgaugetraf}. Connecting such patches results in a globally defined Euler connection for non-degenerate momenta that is unique and gauge invariant.

The Euler loop defined in Eq.~\eqref{eq:Eulerloop} (called the off-diagonal Berry phase in Ref.~\cite{Ahn_2019}) is gauge invariant, since the Euler connection is gauge invariant. 
This means that unlike the Berry phase (i.e.~the integral of the Berry connection) it is not defined only modulo $2\pi$, but can take any real value. It was shown in Ref.~\cite{Ahn_2019} that the winding number $w_i$ around a Dirac point equals the Euler loop $\mathcal{E}_{\ell_i}$ for $\ell_i$ an infinitesimal loop around the $i$-th Dirac point.

For an isolated pair of bands, the sum of all winding numbers $w_i$ is conserved. This means that Dirac points with opposite winding numbers mutually annihilate, while those with equal winding numbers cannot. When the pair of bands is connected to a third band with non-trivial degeneracies between the third band and the pair, this is no longer true. To see this, consider the Dirac string (corresponding to a discontinuity of the eigenvectors in the BZ, given a choice of gauge) connecting Dirac points within a pair of bands~\cite{Ahn_2019}. The shape of the Dirac string within the BZ can be deformed by changing the gauge, but the string cannot be removed. If a change of parameters in the Hamiltonian causes a Dirac point connecting bands one and two to cross a Dirac string associated with bands two and three, this results in the winding number of the Dirac point changing sign. 
The value of individual winding numbers is thus no longer protected, and only the parity of the sum of all winding numbers in any pair of bands is conserved.

In systems where both $\mathcal{T}$ and $\mathcal{C}_2$ are present separately, additional constraints affect the winding numbers of Dirac points that may appear.
Because we assume $\mathcal{T}^2=-1$, there are Kramers degeneracies at all time-reversal invariant momenta (TRIM) within every pair of bands related by $\mathcal{T}$. These degeneracies carry a winding number just like other Dirac points. Moreover, Dirac points not at the TRIM must come in pairs, since for any given Dirac point, the symmetry guarantees a second of the same winding number at the opposite momentum.

The Euler class in systems with both $\mathcal{T}$ and $\mathcal{C}_2$ can thus be expressed as:
\begin{equation}
\label{eq:eulerclass}
    \chi = \frac{1}{2}\sum\limits_{k\in\text{TRIM}} w_k + \sum\limits_{i} w_i.
\end{equation}
Here $i$ runs over the Dirac points in a fundamental domain covering half of the BZ.

\section{Fragility of the Euler class}
\label{methods:fragileEuler}

The identification of the Euler class as a topological invariant is strictly valid only a pair of separated sub-bands. 
In the presence of a third band connected to the two sub-bands, the winding numbers of the Dirac points can flip signs depending on the gauge choice.
This way the total winding number and thus the Euler class become ill-defined, and only the parity of the total winding number will carry physical meaning \cite{Ahn_2019}.
In the presence of time-reversal symmetry, the parity is equivalent to the Fu-Kane-Melee (FKM) invariant.

\begin{figure}[tb]
    \centering
    \includegraphics[width=8.6cm]{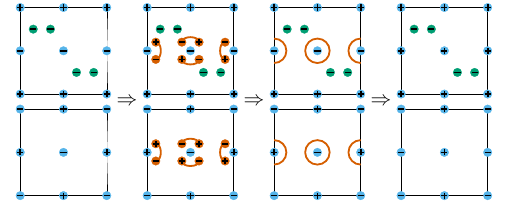}
    \caption{Schematic figure for the change of the Euler class by the addition of a trivial pair of bands. The top row shows the initially non-trivial band with $\chi=-2$. The bottom row shows the added trivial band. Orange points are the degeneracies formed between the topological and trivial pair. At the end we are left with two separated trivial bands with $\chi=0$.
    }
    \label{fig:equivalence}
\end{figure}

As an example, it can be shown that in the model of Eq.~\eqref{eq:Ham}, the $\chi=-2$ phase can be made trivial without closing the gap at the Fermi energy if we allow for an additional pair of occupied bands.
This process is shown schematically in Fig.~\ref{fig:equivalence}.
We first add a trivial pair of bands with $\chi=0$ and create four pairs of Dirac points between the topological and the trivial bands (orange points) around the $\Gamma$ and $X$ points.
Then we annihilate these points such that we form two Dirac string rings around $\Gamma$ and $X$. Contracting the rings will result in changing the winding number at these two TRIM points for both pairs. For the pair of bands that was originally trivial this leaves the Euler class unaffected. For the pair that started out topologically non-trivial, however, it changes the Euler class by 2, resulting in a trivial pair of bands.

With these kind of processes we can change the Euler class of any pair of bands by 2 upon adding a trivial set of bands. Notice that the parity of the Euler class is preserved in this process, so that the FKM invariant is robust even in the presence of multiple bands.

\section{Concentric Wilson loop spectrum}
\label{methods:CWLS}

In general, the calculation of Wilson loops (WL) and Wilson loop spectra (WLS) provides an efficient method to compute topological invariants, in particular in time-reversal symmetric settings. To be specific, we will consider here a system with $2N$ occupied bands (forming $N$ Kramers pairs). For these occupied bands we define the non-Abelian Berry connection matrix $\mathbf{A}_{mn}(\mathbf{k}) = i\mel{u_m(\mathbf{k})}{\nabla_k}{u_n(\mathbf{k})}$ \cite{Yu_2011}. 
The WL matrix is then defined on a closed loop ${\ell}$ as:
\begin{equation}
\label{eq:wilson_methods}
{\cal W}_\ell = {\cal P}\text{ exp}\left(i\oint_{\ell}d\mathbf{k}\cdot\mathbf{A}\right).
\end{equation}
Here ${\cal P}$ is the path ordering operator.

The two Lau-van den Brink-Ortix (LBO) line invariants $\nu_x$ and $\nu_y$ are determined by the eigenvalues of the WL matrix computed along a linear loop $\ell$ in the $k_x$ or $k_y$ direction and across the $\Gamma$ point~\cite{Lau_2016}. For these loops, the combination of $\mathcal{C}_2$ and $\mathcal{T}$ symmetries guarantees that ${\cal W}_\ell=\pm \mathbb{I}$ with $\mathbb{I}$ the identity matrix for any isolated set of two bands. Closing and reopening the gap between such isolated pairs can change the individual WL matrices, but conserves the product of their signs. The signs of ${\cal W}_\ell$ along the $k_x$ and $k_y$ directions thus yield two independent $\mathbb{Z}_2$ invariants~\cite{Lau_2016, Kruthoff_2017}. 

For any closed loop $\ell$, the WL matrix is a unitary matrix whose eigenvalues can be written in the form $\mathrm{e}^{i\varphi_{\ell,n}}$, with $n$ indexing the different eigenvalues. 
Taking a set of loops $\ell_t$ continuously parametrized by $t\in[0,1]$ in such a way that the set covers the Brillouin zone (BZ), and plotting the phases $\varphi_{\ell,n}$ of the WL eigenvalues as a function of $t$, we obtain a WLS.
The FKM invariant $\nu$ can be computed by taking the linear WLS (LWLS) with $\ell_t$ being parallel linear loops across the BZ. For any isolated set of two bands the winding number of the LWLS modulo 2 will yield their FKM invariant~\cite{Kane_2005, Yu_2011}. Closing and reopening gaps between pairs of bands conserves the parity of the sum of windings, and hence the overall FKM invariant.

Here, we additionally consider a set of loops that are concentric around the $\Gamma$ point and respect the $\mathcal{C}_2$ symmetry. Taking the parameter $t$ in $\ell_t$ to be the loop radius, this gives the concentric WLS (CWLS), which has been shown to characterize topological phases in models combining $\mathcal{T}$ and $C_3$ symmetries~\cite{Henke_2021}.

When evaluating the WL matrix of Eq.~\eqref{eq:wilson_methods}, the loop $\ell$ can be divided into $N$ intervals separated by points $\vb{k}_i$, so that the expression of Eq.~\eqref{eq:wilson_methods} can be written as a limit:
\begin{align}\label{eq:Wilsonlimit}
\left(\cal{W}_\ell\right)_{mn} = &\lim\limits_{N\to\infty}\sum\limits_{a_1=1}^2\sum\limits_{a_2=1}^2\dots \sum\limits_{a_N=1}^2 \braket{u_m(\vb{k}_0)}{u_{a_1}(\vb{k}_{1})} \notag \\
&\braket{u_{a_1}(\vb{k}_1)}{u_{a_2}(\vb{k}_{2})}
\braket{u_{a_2}(\vb{k}_2)}{u_{a_3}(\vb{k}_{3})}\dots \notag \\
&\dots
\braket{u_{a_N}(\vb{k}_N)}{u_{n}(\vb{k}_{0})}.
\end{align}
Because the WL matrix is unitary it can be parametrized for two bands as :
\begin{equation}
    \mathcal{W}_\ell = \mathrm{e}^{i\phi_{\ell,0}}\left[\cos{(\phi_{\ell,1})}\sigma_0+i\sin{(\phi_{\ell,1})}\vb{n}\cdot\vb*{\sigma}\right].
\end{equation}
Here $\phi_{\ell,0}$ and $\phi_{\ell,1}$ are two phases, $\vb{n}$ is a unit vector, and $\sigma_0$ and $\vb*{\sigma}$ are the identity and Pauli matrices respectively. For 2D systems exhibiting $\mathcal{C}_2\mathcal{T}$ symmetry with $(\mathcal{C}_2\mathcal{T})^2=1$ (note that only the combined symmetry is required here, not the individual $\mathcal{C}_2$ and $\mathcal{T}$ symmetries), there is always a unitary transformation that renders the Bloch Hamiltonian and its eigenvectors real~\cite{Ahn_2019, Bouhon_2020}. 
In terms of these real eigenvectors, the WL matrix simplifies to:
\begin{equation}
    \mathcal{W}_\ell = \mathrm{e}^{i\phi_{\ell,0}}\left[\cos{(\phi_{\ell,1})}\sigma_0+i\sin{(\phi_{\ell,1})}\sigma_y\right],
\end{equation}
with $\phi_{\ell,0} = 0$ or $\phi_{\ell,0}=\pi$.
The eigenvalues of the WL matrix are given by $\mathrm{e}^{i\varphi_{\ell,\pm}}$ with eigenphases
\begin{equation}
    \varphi_{\ell,\pm} = \phi_{\ell,0}\pm\phi_{\ell,1} \mod{2\pi}.    
\end{equation}

\section{Connection between concentric Euler loops and concentric Wilson loops}
\label{methods:WilsonEuler}

In this section we show how the Wilson loop and the Euler loop are related to each other for real systems with two bands in the presence of $\mathcal{C}_2$ and $\mathcal{T}$ symmetries.
In the real gauge, the eigenstates generally cannot be defined continuously, and discontinuities are found along gauge-dependent Dirac strings. 
Upon traversing a Dirac string, the phase of eigenstates is changed by $\pi$, or equivalently, the eigenstates are multiplied by minus one. Notice that it is always possible to choose a gauge in which the Dirac strings on the two bands making up a Kramers pair coincide throughout the BZ. From here on, we therefore consider only gauge transformations of the form Eq.~\eqref{eq:realgaugetraf} with $\xi_m=\xi~\forall m$, which maintain the coincidence of Dirac strings in pairs of bands.

The non-Abelian Berry connection is not well-defined when traversing a Dirac string, and this should  be taken into consideration when evaluating Eq.~\eqref{eq:wilson_methods}.
A concentric WL always crosses at least one Dirac string, which connects the $\Gamma$ point to one of the other TRIM (with different connections appearing in different gauge choices). The Dirac string connecting the remaining two TRIM can be chosen to lie along the edge of the BZ.
All other Dirac strings will come in pairs, because of the separate $\mathcal{C}_2$ and $\mathcal{T}$ symmetries, which guarantee that each pair of Dirac points connected by a Dirac string has a symmetry related partner. 
Altogether, this implies that every $\mathcal{C}_2$ invariant loop will always cross an odd number of Dirac strings.

\begin{figure}[tb]
    \centering
    \includegraphics[width=8.6cm]{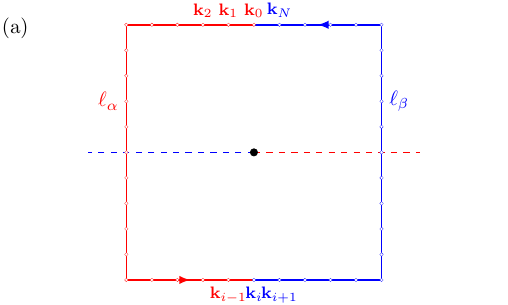}
    \includegraphics[width=8.6cm]{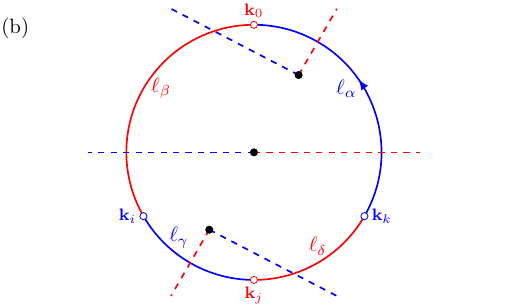}
    \caption{
    \label{fig:looppatch}
    The different gauges used to patch together the Wilson loop in the case of (a) one Dirac string crossing the loop and (b) three Dirac strings crossing the loop.
    The solid red and blue lines indicate the different parts of the Wilson loop, evaluated in different gauges.
    The black dots represent the Dirac points, and dashed red/blue lines indicate the positions of the Dirac strings in the two different gauges, $\alpha$ and $\beta$.
    Notice that these two gauges are chosen such that the respective portion of the Wilson loop only contains continuous eigenvectors, never crossing a Dirac string. That is, the solid blue part of the Wilson loop never crosses a dashed blue Dirac string (and the same holds for red).
    }
\end{figure}

We first discuss the case of a single Dirac string crossing the loop $\ell$, and then generalize to an odd number of crossings.
Even for only a single Dirac string crossing $\ell$, the eigenvectors cannot be made continuous along the entire loop using a single gauge choice.
We therefore divide the loop into two parts $\ell_\alpha$ and $\ell_\beta$ as shown in Fig.~\ref{fig:looppatch}(a). 
We then define two gauges $\ket{u_n^\alpha(\vb{k})}$ and $\ket{u_n^\beta(\vb{k})}$ that are continuous on $\ell_\alpha$ and $\ell_\beta$. 
With these gauges, the WL matrix can be written as:
\begin{align}
\left(\cal{W}_\ell\right)_{mn} = &\lim_{N\to\infty}\sum_{a_1=1}^2\dots \sum_{a_N=1}^2 
\braket{u_m^\alpha(\vb{k}_0)}{u_{a_{1\vphantom{-1}}}^{\alpha\vphantom{\beta}}(\vb{k}_{1})} \dots \notag \\
&\braket{u^\alpha_{a_{1\vphantom{-1}}}(\vb{k}_1)}{u^{\alpha\vphantom{\beta}}_{a_2}(\vb{k}_{2})}\dots
\braket{u_{a_{i-1}}^\alpha(\vb{k}_{i-1})}{u_{a_i}^{\beta}(\vb{k}_{i})} \notag \\
&\dots\braket{u_{a_{N\vphantom{-1}}}^\beta(\vb{k}_N)}{u_{n}^\alpha(\vb{k}_{0})}.
\end{align}

Taking the $N\to\infty$ limit, the WL matrix becomes:
\begin{align}
\label{eq:wilsonsplit}
\left(\cal{W}_\ell\right)_{mn} &= \sum_{a,b,c} \left[\exp\left(i\int\limits_{\ell_{\alpha\vphantom{\beta}}} \vb{dk}\cdot \vb{A}^{\alpha}\right)\right]_{ma} \hspace{-15pt} \braket{u_a^\alpha(\vb{k}_{i})}{u_{b}^{\beta}(\vb{k}_{i})} \notag \\ &\left[\exp\left(i\int\limits_{\ell_\beta} \vb{dk}\cdot \vb{A}^{\beta}\right)\right]_{bc}
\hspace{-10pt}\braket{u_{c}^\beta(\vb{k}_0)}{u_{n}^\alpha(\vb{k}_{0})}.
\end{align}
Here, the path ordering can be neglected, and the integral may be evaluated first and then exponentialized. 
This is possible because of the simple form of the Berry connection in Eq.~\eqref{eq:connection}, for which matrices at different momenta always commute.

Because we consider only real gauges, we know that $\braket{u_a^\alpha(\vb{k})}{u_{b}^{\beta}(\vb{k})}=(-1)^{\xi(\vb{k})}\delta_{ab}$. To define a smooth gauge on both $\ell_\alpha$ and $\ell_\beta$ using a single integer function $\xi(\vb{k})$, it suffices to take a function that changes parity whenever it crosses a Dirac string (in either region). For the situation of Fig.~\ref{fig:looppatch}, a possible choice is to define $\xi(\vb{k})=0$ on the upper half of the BZ, and $\xi(\vb{k})=1$ on the lower half.
Since the Euler connection in Eq.~\eqref{eq:connection} is invariant under gauge transformations of the form of Eq.~\eqref{eq:realgaugetraf}, it can be defined across the entire loop $\ell$, and the integrals in Eq.~\eqref{eq:wilsonsplit} can be united to yield:
\begin{align}
\mathcal{W}_\ell &= (-1)^{\left[\xi(\vb{k_i})+\xi(\vb{k_0})\right]}\exp\left(i\oint\limits_{\ell=\ell_\alpha+\ell_\beta} \vb{dk}\cdot \vb{A}\right) \notag \\
&=-\exp\left(-i\sigma_y\oint\limits_{\ell} \vb{dk}\cdot \vb{a}(\vb{k})\right)=-\mathrm{e}^{-i\pi \sigma_y\mathcal{E}_\ell} \notag \\
&=-\left[\cos{(\pi\mathcal{E}_\ell)}\sigma_0+i\sin{(-\pi\mathcal{E}_\ell)}\sigma_y\right].
\end{align}
This expression shows that the phases of the eigenvalues of the Wilson loop matrix are related to the Euler loop along the same $\mathcal{C}_2$ symmetric path by:
\begin{equation}
\label{eq:wilsoneignew}
    \varphi_{\ell,\pm} = \pi(1 \pm \mathcal{E}_\ell) \mod 2\pi.
\end{equation}

This relation between Wilson and Euler loops along $\mathcal{C}_2$ symmetric paths can be extended to loops crossing multiple Dirac strings.
An example for this is shown in Fig.~\ref{fig:looppatch}(b).
Again, we can define two gauges and divide the loop $\ell$ into multiple pieces (the four lines $\ell_\alpha$, $\ell_\beta$, $\ell_\gamma$, and $\ell_\delta$ in the example of Fig.~\ref{fig:looppatch}(b).
Following similar arguments as in the case of a single Dirac string, the WL matrix can be written as:
\begin{align}
\mathcal{W}_\ell =& (-1)^{\left[\xi(\vb{k}_i)+\xi(\vb{k}_j)+\xi(\vb{k}_k)+\xi(\vb{k}_0)\right]} \notag \\
&\exp\left(i\oint\limits_{\ell=\ell_\alpha+\ell_\beta+\ell_\gamma+\ell_\delta} \vb{dk}\cdot \vb{A}\right).
\end{align}
Here, $\xi(\vb{k})$ is an integer function that changes parity across every Dirac string (both red and blue). 
Using this, it follows that for an even (odd) number of Dirac strings crossing the loop $\ell$ in a given gauge, the overall sign of the WL matrix is positive (negative). As discussed above, in systems with $\mathcal{C}_2$ and $\mathcal{T}$ symmetries, there is always an odd number of Dirac strings crossing any $\mathcal{C}_2$ symmetric contour, and the Wilson loop will therefore always be of the form $\mathcal{W}_\ell=-\mathrm{e}^{-i\pi\sigma_y\mathcal{E}_\ell}$, and its eigenphases are given by Eq.~\eqref{eq:wilsoneignew}.

\section{Density functional theory calculations}
\label{methods:DFT}

All density functional theory (DFT) calculations were done using the full-potential local-orbital code FPLO~\cite{Koepernik_99} version 22. For the exchange and correlation potential, we used the generalized gradient approximation~\cite{Perdew_96}. All calculations are done on a $k$-mesh of $n\times{}m\times{}1$ points, where $n$ and $m$ are chosen such that the grid density is $\sim5000$ $k$-points per reciprocal atom and the grid is uniformly spaced (see Table~\ref{tab:dft}). As an input for DFT calculations, we used structures from C2DB~\cite{Haastrup_2018} that were subsequently symmetrized using FINDSYM version 7.1.3~\cite{Stokes_05}. Full relativistic self-consistent GGA calculations were followed by a direct construction of Wannierized Hamiltonians describing the entire valence band, which is done by L\"owdin-orthogonalization of the local-orbital basis. Crucially, the procedure~\cite{Koepernik_2023} implemented in FPLO ensures that resulting tight-binding Hamiltonians in the Wannier basis retain time-reversal and all space-group symmetries.

\begin{table}[tb]
\caption{\label{tab:dft} Space groups and $k$-meshes used for DFT calculations of five 2D materials.}
\begin{ruledtabular}
\begin{tabular}{rrrr}
material             & Ref.              & space group  & $k$-mesh \\ \hline
Au$_2$SO$_4$         & \cite{Au2SO4}     & $C222$       & $27\times{}27\times1$ \\
Au$_3$InO$_4$Br$_4$  & \cite{Au3InO4Br4} & $P2$         & $21\times{}20\times1$ \\
MoOClBr              & \cite{MoOBrCl}    & $Cmm2$       & $25\times{}25\times1$ \\
MoAg$_2$Te$_4$       & \cite{MoAg2Te4}   & $P\bar{4}2m$ & $27\times{}27\times1$ \\
PdIr$_3$S$_4$Br$_4$  & \cite{PdIr3S4Br4} & $Cmm2$       & $20\times{}20\times1$ \\
\end{tabular}
\end{ruledtabular}
\end{table}

\section{Additional material results}
\label{methods:materials}

Here we summarize our results for all five materials.
In each case, we plot both the gap as a function of momentum, and the CWLS and Euler loops, as done in Fig.~\ref{fig:MoAg2Te4} of the main text.
Since the materials belong to different symmetry groups, in each case we parameterize the BZ using two dimensionless momenta, $k_{1,2}={\bf k}\cdot {\bf a}_{1,2}$, where ${\bf a}_{1,2}$ are the two Bravais vectors.
In those cases where the latter are not perpendicular, this corresponds to a distortion of the BZ, which, however, still fully encodes all available information.
We remark that the concentric loops are in all cases defined in this (possibly distorted) $k_{1,2}$ space, which guarantees that the loops obey $\ctwo$ symmetry and all of our results apply.

For MoAg$_2$Te$_4$, in addition to the results shown in Fig.~\ref{fig:MoAg2Te4} for the third band pair, we find that generic Dirac cones enforced by fragile topology exist in band pairs 17, 29, and 30. Out of these, the band pair 29 is closest to the Fermi level, positioned between $0.69$ eV and $1.07$ eV above $E_F$.

\begin{figure*}[tb]
    \centering
    \includegraphics[width=0.85\linewidth]{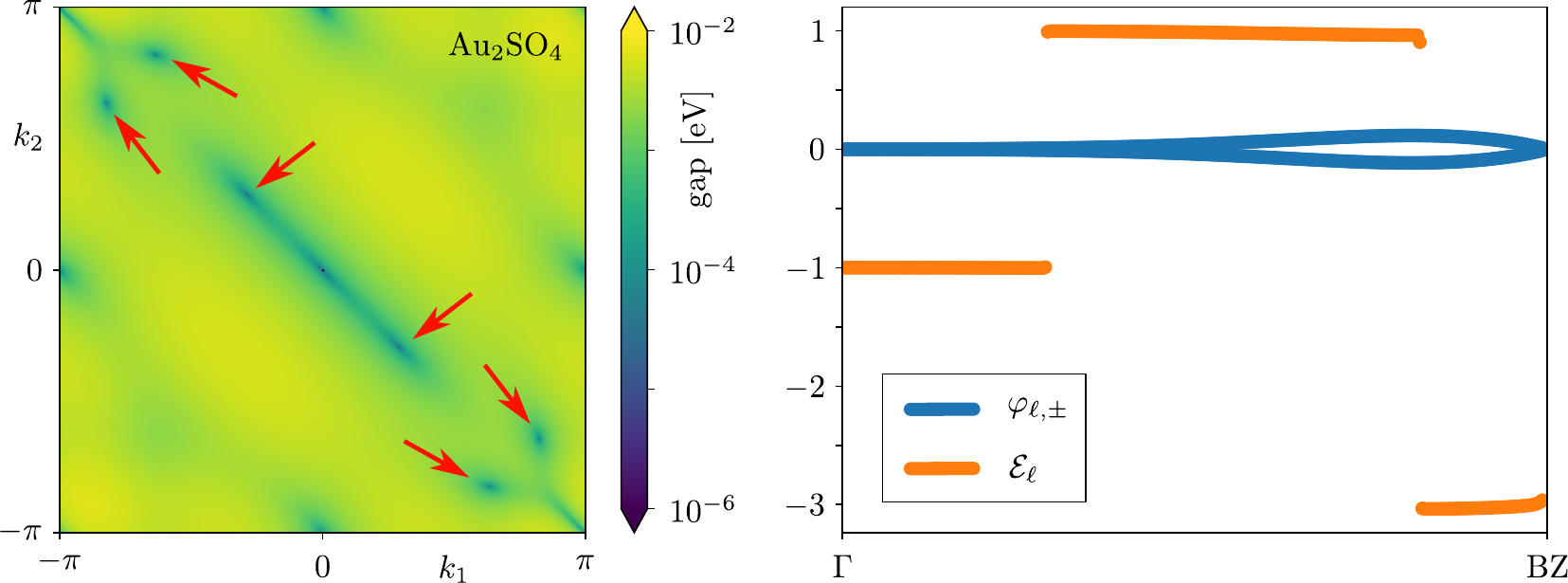}
    \caption{Dirac cones enforced by fragile topology in Au$_2$SO$_4$, using the same conventions as Fig.~\ref{fig:MoAg2Te4}.
    Generic Dirac cones are marked with red arrows.
    The two cones close to $\Gamma$ have winding numbers $+1$, and the four cones further away have winding numbers $-1$, as can be seen from the Euler loop calculation in the right panel.
    This is band pair number 8, with energies between $-8.2$ eV to $-7.9$ eV (below $E_F$).}
    \label{fig:Au2SO4}
\end{figure*}

For Au$_2$SO$_4$, $Q\neq 0$ occurs in band pairs 3, 4, 8, 93, and 94.
Results for band pair 8 are shown in Fig.~\ref{fig:Au2SO4}.
The sum of TRIM cone winding numbers is $+2$ and the CWLS does not wind, yielding $Q=-1$.
Note that according to our results this index predicts at least two generic Dirac cones, whereas the band structure shows six in total, four of which come in pairs with opposite winding numbers.

\begin{figure*}[tb]
    \centering
    \includegraphics[width=0.85\linewidth]{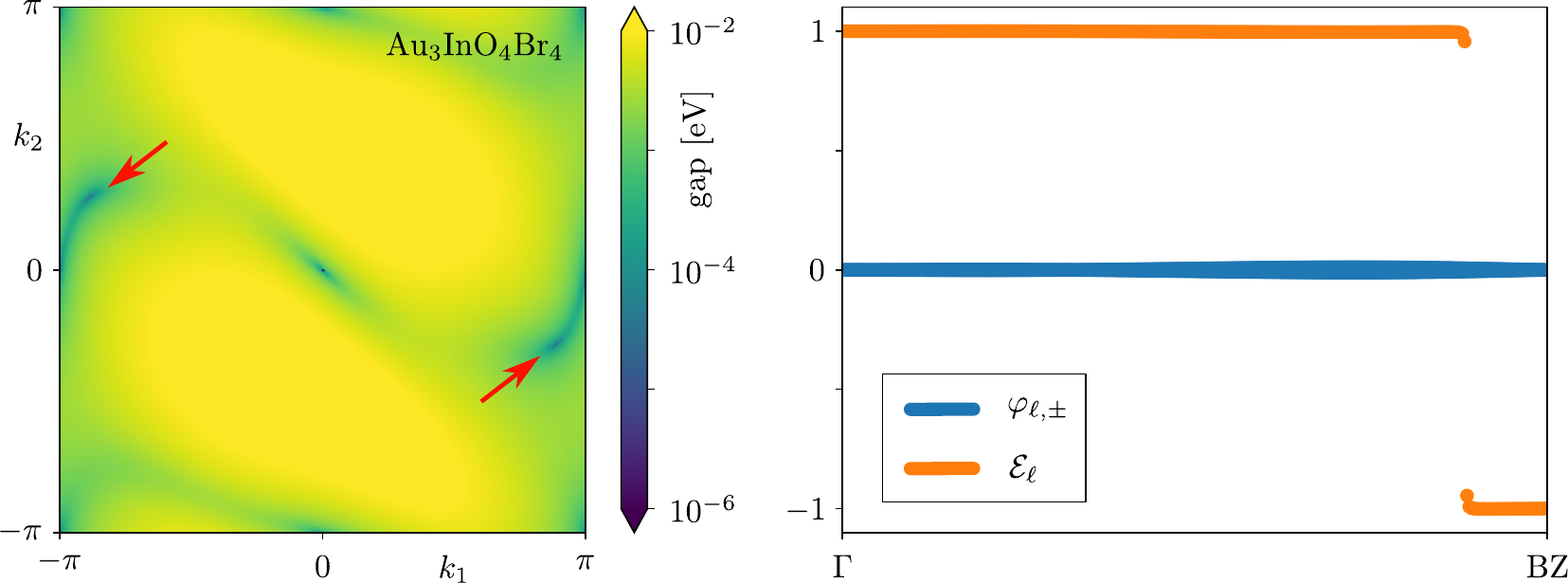}
    \caption{Dirac cones enforced by fragile topology in Au$_3$InO$_4$Br$_4$, using the same conventions as Fig.~\ref{fig:MoAg2Te4}.
    Both generic Dirac cones are marked with red arrows and have winding numbers $-1$.
    This is band pair number 7, with energies around $-14.8$ eV (deep in the valence sector).}
    \label{fig:Au3InO4Br4}
\end{figure*}

For Au$_3$InO$_4$Br$_4$, $Q\neq 0$ in band pairs 7, 53, 54, 147, and 167.
We show results for band pair number 7 in Fig.~\ref{fig:Au3InO4Br4}.
Here again, the sum of TRIM cone winding numbers is $+2$ and there is no winding of the CWLS, such that $Q=-1$.
The minimum number of two generic Dirac cones is realized in the band pair, as indicated by the red arrows.

\begin{figure*}[tb]
    \centering
    \includegraphics[width=0.85\linewidth]{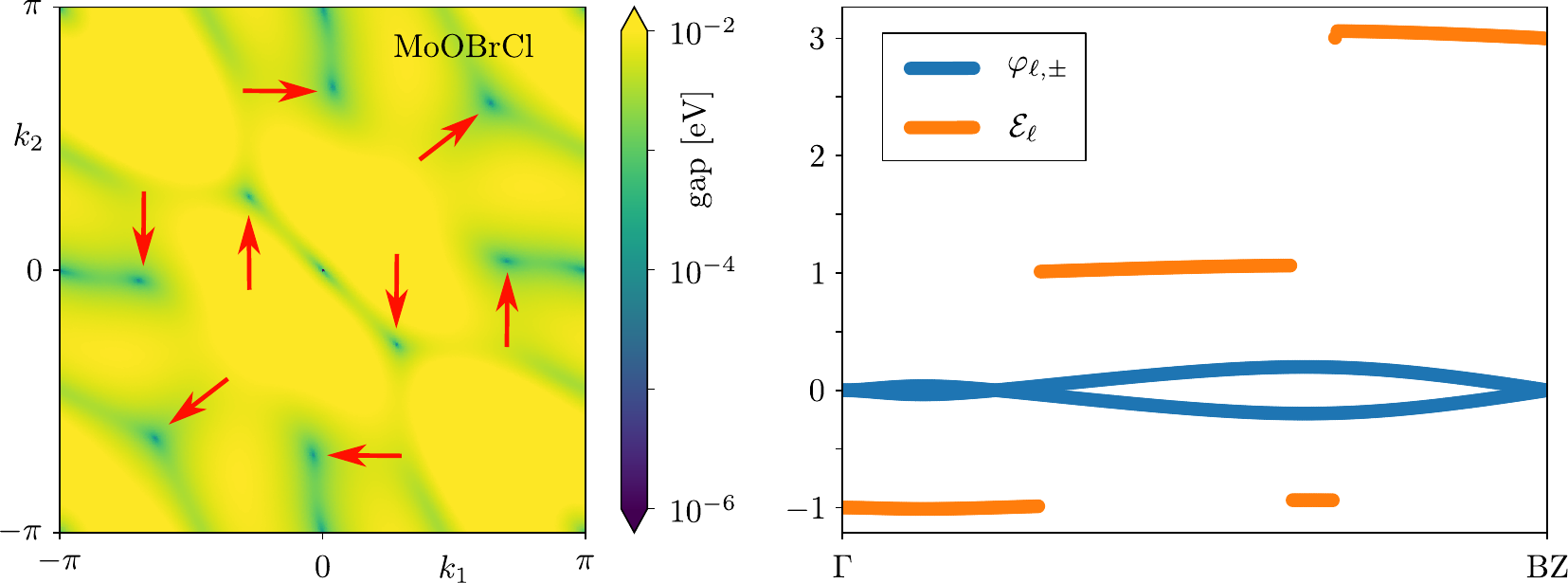}
    \caption{Dirac cones enforced by fragile topology in MoOBrCl, using the same conventions as Fig.~\ref{fig:MoAg2Te4}.
    Generic Dirac cones are marked with red arrows.
    This is band pair number 25, with energies between $-1.94$ eV to $-1.74$ eV (below $E_F$).}
    \label{fig:MoOBrCl}
\end{figure*}

In MoOBrCl, we find that band pairs 3, 4, 5, 6, 24, and 25 have $Q\neq 0$. Results for the latter pair are shown in Fig.~\ref{fig:MoOBrCl}, showing a total of 8 generic Dirac cones.
The TRIM cone winding numbers sum to $-4$ and the CWLS does not wind, leading to $Q = 0 - \left| \frac{-4}{2} \right| = -2$ according to Eq.~\eqref{eq:Q} of the main text.
Correspondingly, there are four generic Dirac cones that all have winding numbers $+1$ (those located closest to $k_1=0$ or to $k_2=0$), while the remaining four generic cones come in pairs with winding numbers $\pm 1$.
Note that the sign of $Q$ does not predict the sign of the generic cone winding numbers, given the absolute value present in the second term of Eq.~\eqref{eq:Q}.
Instead, the index predicts the minimum number of generic cones, which in this case is $2|Q|=4$.
In addition, we remark that this band pair is an example highlighting the difference between the Euler class and the fragile index $Q$.
The sum over all winding numbers of all Dirac cones (TRIM and generic) vanishes, $\chi=0$, but in spite of this the band pair is not trivial.

Except for the degeneracy points, the two bands in this pair have a separation in energy around 10 meV or even higher.
In addition, the band is positioned between $-1.94$ eV to $-1.74$ eV below $E_F$, and is well separated from adjacent band pairs (167 meV to the pair below and 687 meV to the pair above).
The combination of these features makes it a potentially good candidate for ARPES experiments.

\begin{figure*}[tb]
    \centering
    \includegraphics[width=0.85\linewidth]{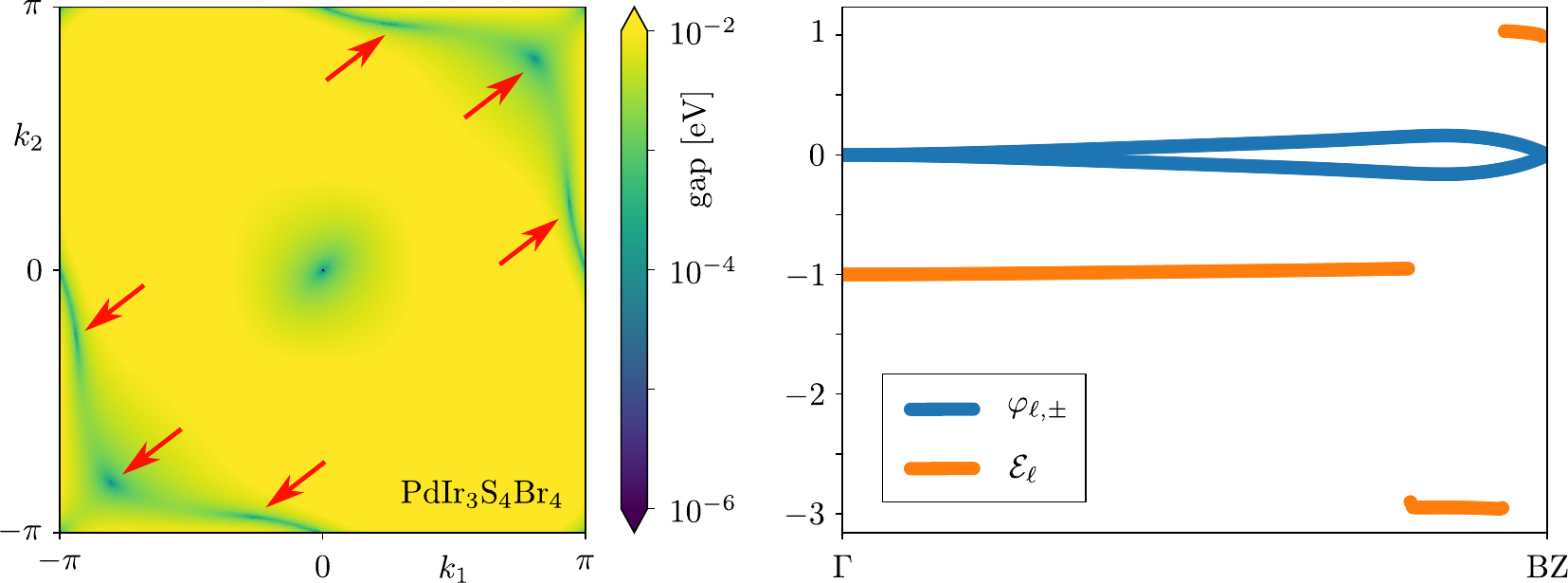}
    \caption{Dirac cones enforced by fragile topology in PdIr$_3$S$_4$Br$_4$, using the same conventions as Fig.~\ref{fig:MoAg2Te4}.
    Generic Dirac cones are marked with red arrows.
    This is band pair number 8, with energies around $-14$ eV (deep in the valence sector).}
    \label{fig:PdIr3S4Br4}
\end{figure*}

Finally, PdIr$_3$S$_4$Br$_4$ shows $Q\neq 0$ in band pairs 7, 8, and 56. 
Band pair number 8 is shown in Fig.~\ref{fig:PdIr3S4Br4}, for which $Q=1$.
Two generic cones have winding numbers $+1$, and the other four come in pairs of winding numbers $\pm 1$.

The results for all other band pairs which are not shown here can be reproduced using the code available at \cite{Zenodo_code}.
Note that in this python code the numbering of pairs starts from zero instead of one.

\clearpage
\bibliography{references}

\end{document}